# Many-Electron Effects on Optical Absorption Spectra of Strained Graphene

Yufeng Liang, Shouting Huang and Li Yang*

Department of Physics, Washington University, St. Louis, Missouri 63130, USA
(* corresponding author)

## ABSTRACT

We employ the first-principles GW+Bethe Salpeter equation approach to study the electronic structure and optical absorption spectra of uniaxial strained graphene with many-electron effects included. Applied strain not only induces an anisotropic Fermi velocity but also tilts the axis of the Dirac cone. As a result, the optical response of strained graphene is dramatically changed; the optical absorption is anisotropic, strongly depending on the polarization direction of the incident light and the strain orientation; the characteristic single optical absorption peak from π-π* transitions of pristine graphene is split into two peaks and both display enhanced excitonic effects. Within the infrared regime, the optical absorbance of uniaxial strained graphene is no longer a constant because of the broken symmetry and associated anisotropic excitonic effects. Within the visible-light regime, we observe a prominent optical absorption peak due to a significant red shift by electron-hole interactions, enabling us to change the visible color and transparency of stretched graphene. Finally, we also reveal enhanced excitonic effects within the ultraviolet regime (8 to 15 eV), where a few nearly bound excitons are identified.

## I. INTRODUCTION

Graphene, a single layer of graphite,[1-4] has ignited tremendous research attention because of interesting physics associated with its unusual electronic structure and promising device applications.[5-6] In particular, the optical response of graphene displays many intriguing features, such as a nearly constant optical conductivity in the infrared regime and gate-dependent optical absorbance.[7-11] Meanwhile, graphene exhibits outstanding mechanical properties,[12, 13] for example, it can sustain a huge uniaxial stretch (up to 20%),[12] placing graphene among the hardest materials known. This impressive structural modulation through applying uniaxial strain gives hope to a useful way to tailor the electrical and optical properties of graphene for broad applications.

To date, in addition to experimental advances, both model and density functional theory (DFT) calculations have revealed numerous strain effects on the electronic structure and optical response of graphene.[14-17] However, in order to thoroughly understand its optical excitations, many-electron effects, such as electron-electron (*e-e*) and electron-hole (*e-h*) interactions, have to be included in a more accurate way because they are known to be important factors in deciding excited properties of both pristine and doped graphene.[9, 18-20] In particular, since applied strain will inevitably change the hexagonal geometry, it is of fundamental interest to study *e-h* pairs (excitons) and their optical activities under such an anisotropic environment, which have not been well understood yet.

In this work, we apply the first-principles GW+Bethe Salpeter equation (BSE) approach to study quasiparticle energy and optical excitations of uniaxial strained graphene. Two typical stretching directions are considered, *i.e.*, zigzag and armchair cases, respectively. Our simulation shows that applied strain shifts the position of Dirac points and modifies the shape of the Dirac cone, *e.g.*, the Fermi velocity is anisotropic and the Dirac cone is no longer located at corners of the hexagon of the first Brillouin zone (BZ). Moreover, the axis of the Dirac cone is tilted as well, breaking the known $\pi$ and $\pi^*$ symmetry of pristine graphene. At the same time, the degeneracy of important saddle points ($M$ and $M'$ points) is broken, giving rise to double-peak absorption spectra.

The above changes of the electronic structure induce significant variations of the optical response of strained graphene. First, the optical absorption is strongly anisotropic; depending on the polarization direction of the incident light and strain orientation, we can obtain a double-peak or single-peak absorption spectrum. Second, *e-h* interactions are enhanced because applied strain results in flatter band dispersion around those van Hove singularities and increases the effective mass of involved electrons and holes. One particularly useful excitonic effect is that an absorption peak is significantly shifted from the ultraviolet regime into the visible-light regime, making graphene more suitable for optical devices. Third, the infrared optical absorbance is no longer a constant; the broken $\pi$ and $\pi^*$ symmetry and enhanced *e-h* interactions contributed to a varying absorbance. Finally, nearly bound excitons are identified within the higher-frequency regime (8 to 15 eV), which is useful for graphene devices working within the ultraviolet field.

## II. THEORY AND METHODOLOGY

In this work, we follow the widely used approach to study many-electron effects in solids in three steps:[21] (i) we obtain the electronic ground state using DFT within the local density approximation (LDA);[22, 23] (ii) the quasiparticle excitations are calculated within the GW approximation;[24, 25] and (iii) we solve the BSE to obtain the photo-excited states and optical absorption spectra.[18, 21, 26]

We use Troullier-Martins norm-conserving pseudopotentials[27] by employing a plane-wave basis with a 60-Ry energy cutoff. A 29 x 32 x 1 or 32 x 29 x 1 k-grid is used to obtain the converged Kohn-Sham eigenvalues and wave functions for zigzag or armchair strained graphene, respectively. We compute the quasiparticle energy correction by the $G_0W_0$ approximation,[25] in which the energy cutoff of the dielectric function is set to be 8.0 Ry and the static dielectric function is extended to the dynamic case with the generalized plasmon-pole model.[25] A 59 x 64 x 1 or 64 x 59 x 1 k-point sampling is necessary to calculate the converged quasiparticle energy of zigzag or armchair strained graphene, respectively. The optical absorption spectrum with excitonic effects included is obtained by solving the BSE

$$(E_c - E_v)A^s_{vck} + \sum_{v'c'} K^{AA}_{vck,v'c'k}(\Omega_S)A^s_{v'c'k} = \Omega_S A^S_{vck}, \tag{1}$$

with the static *e-h* interaction approximation.[21, 26] In Eq (1), $A^S_{vck}$ is the exciton amplitude, $k$ is the wave vector, $E_c$ and $E_v$ are quasiparticle energy of conducting and valence electrons, respectively. The single-particle (without *e-h* interaction included) and two-particle (with *e-h* interaction included) optical absorption spectra can be calculated by following formulas listed in Ref. 21.

To obtain a converged absorption spectrum up to 15 eV, we consider 2 valence and 7 conduction bands and adopt a 118 x 128 x 1 or 128 x 118 x 1 k-grid on the first BZ of zigzag or armchair strained graphene, respectively. Because of the depolarization effect, we only consider the optical absorption spectra with the incident light polarized parallel to the graphene sheet. In addition, because the above simulation approach does not included intraband transitions, we neglect the optical spectra within the frequency regime less than 0.5 eV, where the Drude effect under a finite temperature may be of importance.[10]

## III. SIMULATION RESULTS AND DISCUSSIONS

In this work, we consider two typical uniaxial stretching directions, zigzag and armchair. The stretching is set to be +10%, which is large enough to emphasize strain effects but still well accessible according to published experimental results.[12] For smaller strain, we believe the interesting physics found in this work is still true but may be in smaller magnitudes. The structure of strained graphene is fully relaxed according to the force and stress (except for the stretching direction) within DFT/LDA. The distance between adjacent graphene layers is set to be 1.5 nm to avoid artificial inter-layer interactions.

### A. Electronic structure of strained graphene

The LDA calculated band structure and the first BZ of zigzag and armchair stretched graphene are shown in Figure 1, respectively. The LDA and GW calculated Fermi velocities are also concluded in Table I. According to Figures 1 (a) and (b), the first BZ is distorted by applied strain and no longer in a perfect hexagonal shape. At the same time, as revealed by previous studies,[15] Dirac points are shifted from corners. In Figure 1 (a), the Dirac points around the $K$ point are shifted out from the first BZ while the other Dirac point around the $K'$ point is shifted into the first BZ along the $K'-M'$ direction for the zigzag strained graphene. In the armchair stretched graphene shown in Figure 1 (b), the shift of Dirac points is the inverse case; the Dirac point around the $K$ point is shifted into the First BZ while the other around the $K'$ point is shifted out. We also notice that this shifting effect is more significant in zigzag stretched graphene but smaller in the armchair stretched case.

In addition to the shift of Dirac points, the Fermi velocity in strained graphene is significantly modified from that of pristine graphene. For example, the LDA-calculated Fermi velocity along the armchair direction of armchair stretched graphene is around 6.9 x $10^5$ m/s as listed in Table I, which is around 20% less than that of pristine graphene (8.5 x $10^5$ m/s). Moreover, this modification of the Fermi velocity is anisotropic. According to Table I, Fermi velocities along the zigzag direction are smaller than that along the armchair directions in the zigzag strained graphene. In particular, *the Fermi velocity along the stretched direction is always smaller than those along the perpendicular direction*. In a word, free carriers shall move faster perpendicularly to the stretching direction for all our studied graphene.

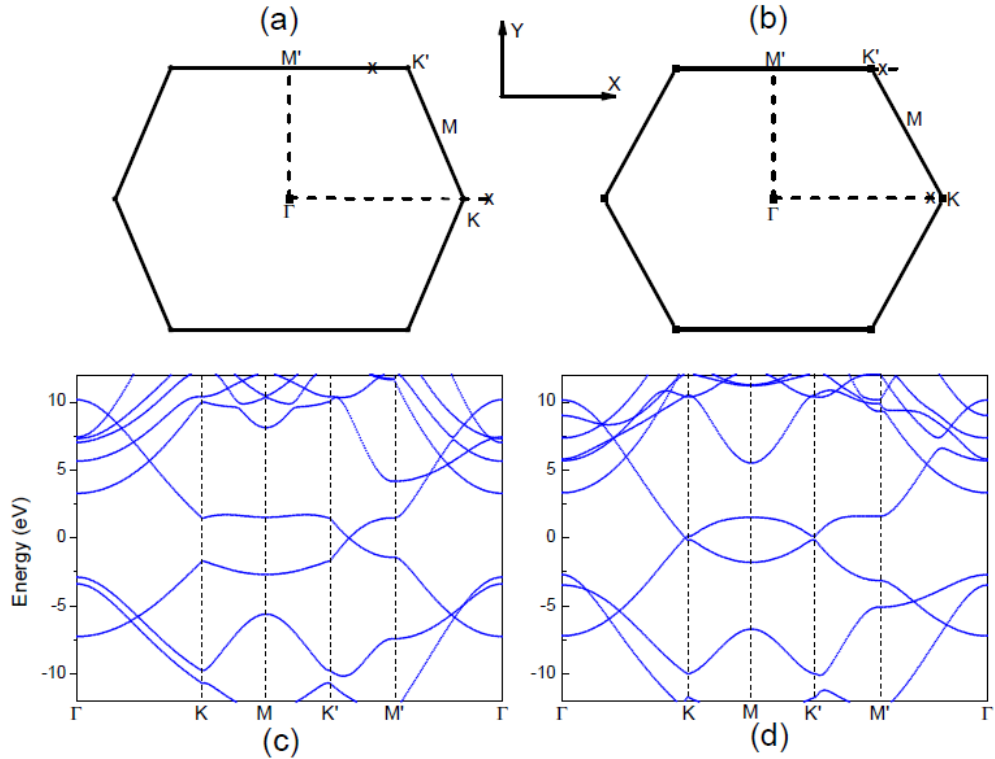

Figure 1. (a) and (b) are the schematic first BZ of zigzag and armchair strained graphene, respectively. Locations of Dirac points are marked by cross symbols. The x axis is along the zigzag direction and the y axis is along the armchair direction. (c) and (d) are the LDA calculated band structure of zigzag and armchair strained graphene, respectively. The energy level of the Dirac point is set to be zero.

Furthermore, as shown in Table I, we find two Fermi velocities along the same direction in stretched graphene. For example, zigzag stretched graphene have two different LDA calculated Fermi velocities along the zigzag direction, 6.4 x $10^5$ m/s and 7.5 x $10^5$ m/s, respectively. This means that the axis of the Dirac cone is tilted. Interestingly, all our calculations show that the tilting direction is always toward the zigzag direction despite the strain direction. This tilted Dirac cone breaks the known symmetry of electrons and holes, the foundation of many important characters of graphene,[7,8] *e.g.*, the constant infrared optical conductivity. Therefore, we expect the strained graphene may exhibit unusual optical responses within the low-frequency excitation regime.

Table I. LDA and GW calculated Fermi velocities of uniaxial stretched graphene. The unit of the velocity is $10^5$ m/s. The x and y direction are defined in Fig. 1.

|     | Zigzag Stretching | | | Armchair Stretching | | |
| --- | --- | --- | --- | --- | --- | --- |
|     | $v_x$ | | $v_y$ | $v_x$ | | $v_y$ |
| LDA | 6.4 | 7.5 | 8.8 | 8.4 | 9.3 | 6.9 |
| GW  | 9.0 | 10.4 | 12.0 | 11.8 | 13.0 | 9.6 |

Other than changes around the Dirac cone, many more strain effects can be identified in Figure 1 (c) and (d). One of them is the broken energy degeneracy at $M$ and $M'$ points. Since these two points are important saddle points related to the van Hove singularity that usually contributes to absorption peaks,[18, 19] such a split gives rise to a double-peak feature on the optical absorption spectrum of strained graphene.

Finally, from table I, we see enhanced self-energy corrections in all stretched graphene studied. For example, the Fermi velocity along the armchair direction of armchair stretched graphene is increased from 6.9 x $10^5$ m/s to 9.6 x $10^6$ m/s, a 40% enhancement which is larger than that of pristine graphene (around 30%).[18]

**B. Optical absorption spectra of zigzag strained graphene**

The optical absorption spectra of pristine and zigzag strained graphene are presented in Figure 2 by blue curves. The most prominent feature of strained graphene is the anisotropic optical absorbance resulted from the broken hexagonal symmetry; only one absorption peak is observed for the incident light polarized along the stretching direction as shown in Figure 2 (b) while two peaks are observed for the incident light polarized perpendicularly to the stretching direction as shown in Figure 2 (c). This polarization dependent absorption is decided by the symmetry of wave functions at $M$ and $M'$ points, respectively, because interband transitions at the $M$ point are active for both zigzag and armchair polarizations while those at the $M'$ point are more active for the armchair polarization.

In addition, we also observe a significant anisotropic effect on the infrared optical absorbance, which is no longer to be around 2.3%,[8] the value of pristine graphene. Although the infrared absorbance of strained graphene calculated by interband transitions is still nearly constant, its value strongly depends on the polarization direction of the incident light. As read out from Figure 2 (b) and (c), the single-particle infrared optical absorbance is around 1.9% for the zigzag polarization while it is 3.0% for the armchair polarization. This is attributed to the anisotropic character of the Dirac cone in strained graphene.

When *e-h* interactions are included, enhanced excitonic effects are exhibited as shown by red curves in Figure 2. First, a significant red shift is observed in all absorption spectra. In figure 2 (b), for zigzag polarized incident light, the prominent optical absorption peak C is shifted from 5.5 eV to 4.4 eV by *e-h* interactions. This 1.1 eV red shift is much larger than that of pristine graphene (~ 600 meV).[18] Such enhanced excitonic effects can be understood from the change of effective mass of involved electrons and holes. Following this idea, we turn to the band structure at the $M$ point plotted in Figure 1 (c) because the absorption peak C is originated from interband transitions there. Since applied strain makes the curvature of the π and π* electronic bands around the $M$ point flatter than those of pristine graphene, larger effective mass is expected for both electrons and holes. According to the hydrogenic model (the effective-mass model), and corresponding excitonic effects shall be enhanced consequently.

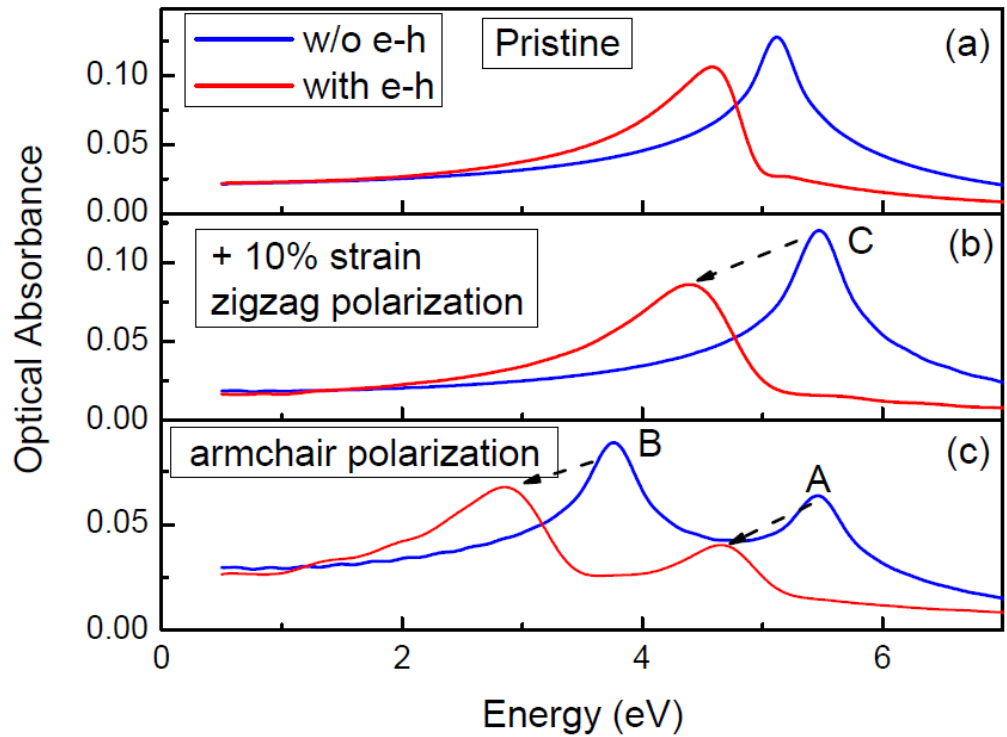

Figure 2. Optical absorption spectra of pristine (a) and zigzag strained graphene with the incident light polarized along the zigzag direction (b) and armchair direction (c). A 0.15 eV Gaussian broadening is applied to obtain the smooth absorption curve. For pristine graphene plotted in (a), the absorption is independent of the polarization direction of incident light.

We also notice that the red shift induced by excitonic effects is different for each absorption peak. For example, the red shift of the absorption peak B in Figure 2 (c) is around 0.9 eV, which is larger that that of the absorption peak A (0.8 eV). This can also be understood by different effective mass associated with $M$ and $M'$ points as shown in Figure 1 (c). The absorption peak B is of particular interest because *e-h* interactions drag it from the ultraviolet regime to the visible-light regime. As a result, we expect significant changes of transparency and color of such strained graphene. Moreover, since the absorption peak B is only observed for the armchair polarization, this polarization dependence of the optical response of zigzag strained graphene can be useful for optical filters working within the visible-light frequency.

Finally, in both Figure 2 (b) and (c), we observe the infrared optical absorbance is qualitatively different from that of pristine graphene; it is no longer a constant optical absorbance after *e-h* interactions are included. This is because the *e-h* symmetry is broken by applied strain as we discuss in section A. In addition, the anisotropic Fermi velocity around the Dirac point induces anisotropic excitonic effects, which further enhances the variation of the infrared optical absorbance.

**C. Optical absorption spectra of armchair strained graphene**

The optical absorption spectra of armchair strained graphene are presented in Figure 3, which shows similar characters as those in Figure 2 of the zigzag strained case; spectra exhibit substantially anisotropic behaviors; enhanced *e-h* interactions result in significant red shift of prominent absorption peaks; the infrared optical absorbance is depending on the direction of the polarization of the incident light and no longer a constant after *e-h* interactions are included. On the other hand, one special feature of this armchair strained graphene is the relative intensity of its two prominent absorption peaks in Figure 3 (c); that of the peak A is higher than that of the peak B. This is the inverse of spectra of zigzag strained graphene shown in Figure 2 (c).

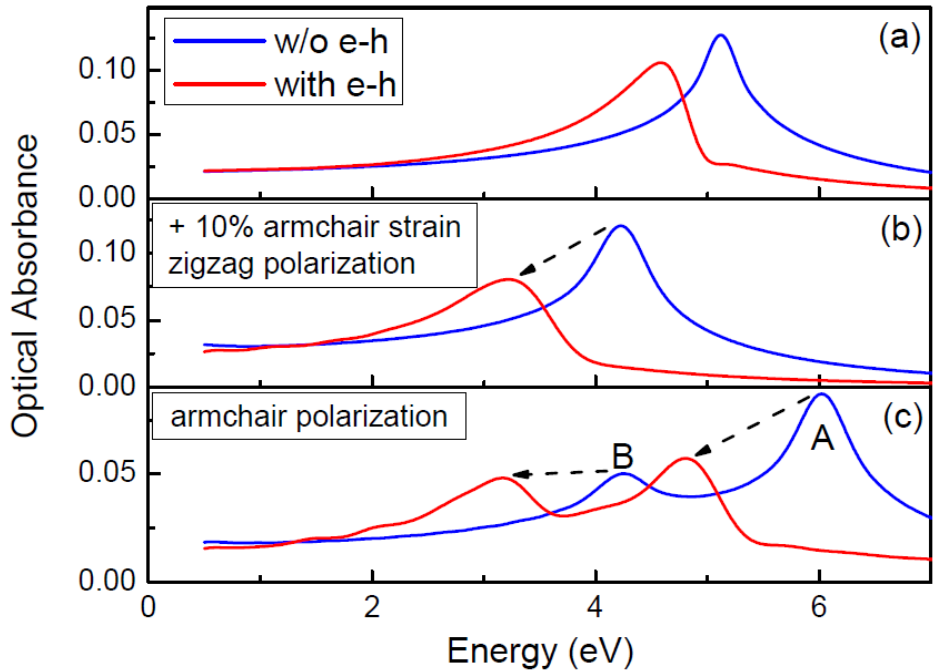

Figure 3. Optical absorption spectra of pristine (a) and armchair strained graphene with the incident light polarized along the zigzag direction (b) and armchair direction (c). A 0.15 eV Gaussian broadening is applied to obtain the smooth absorption curve.

Another useful feature of armchair strained graphene is that its optical absorbance has a peak (B) in the visible-light regime (around 3 eV) for both armchair and zigzag polarizations as shown in Figure 3 (b) and (c). This hints us the armchair strained graphene shall display stronger changes on its transparency and colors than the zigzag strained case.

**D. High-frequency optical absorption spectra of strained graphene**

All optical absorption spectra discussed above are originated from optical activities of π and π* electronic states. Because of the gap-less feature of graphene, resonant excitonic effects are dominant many-electron effects there. When moving to electronic bands far away from the Dirac point, we find σ states are also active in the high-frequency optical absorption spectrum. Enhanced excitonic effects[28] and even nearly bound excitons have been identified[29] in recent *ab initio* calculations. In this section, we will study the optical response and corresponding excitonic effects within such a high-frequency regime of strained graphene.

Presented in Figure 4 are the high-frequency optical absorption spectra of zigzag strained graphene. Because of the broken geometry, the optical absorption spectra of strained graphene are substantially different from that of the pristine case even without *e-h* interaction included. For example, the single-particle optical absorption shoulders (10.9 eV for the zigzag polarization and 11.5 eV for the armchair polarization) of zigzag strained graphene are significantly lower than that (12.4 eV) of pristine graphene.

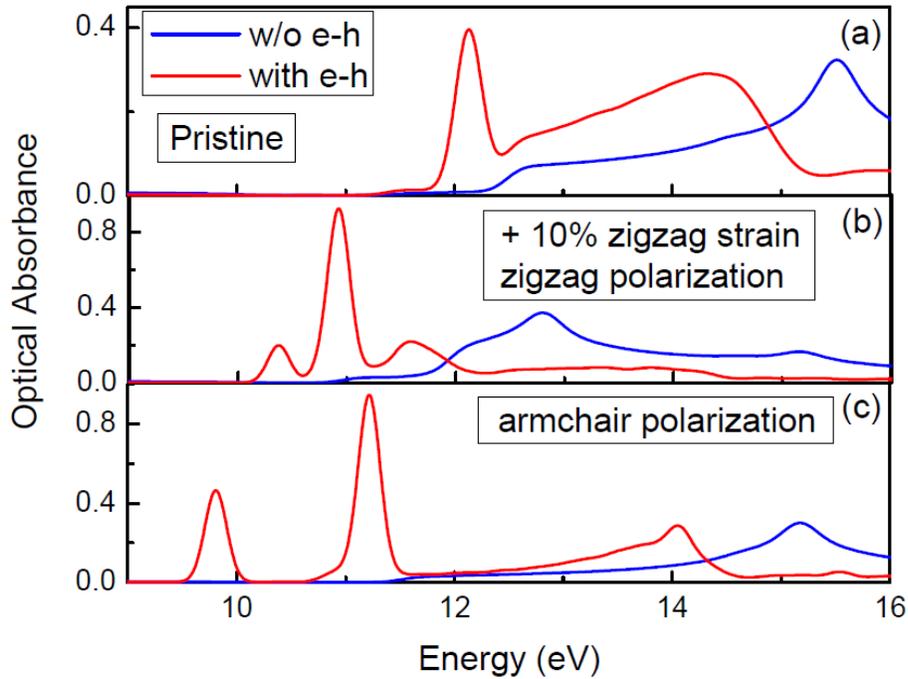

Figure 4. High-frequency optical absorption spectra of pristine (a) and zigzag strained graphene with the incident light polarized along the zigzag direction (b) and armchair direction (c). A 0.15 eV Gaussian broadening is applied to obtain the smooth

absorption curve.

After *e-h* interactions are included, enhanced excitonic effects contributed to a few nearly bound excitons, those isolated absorption peaks shown in Figure 4 (b) and (c). Due to their narrow line width, these nearly bound excitons usually own a much longer lifetime than resonant excitonic states and could be very useful for optoelectronic applications. Interestingly, the binding energy of these excitons calculated in strained graphene in Figure 4 is around a few hundreds meV to 1 eV, which is even larger than that of pristine graphene (around 270 meV). Since the applied strain does not change the semimetallic nature of graphene, we estimate that the screening between electrons and holes shall not be changed too much. On the other hand, we notice that σ states around the Γ point, which significantly contribute to the oscillator strength of the high-frequency optical absorption, have highly anisotropic band dispersion as shown in figure 1 (c) and (d). This highly anisotropic effective mass may enhance corresponding excitonic effects.

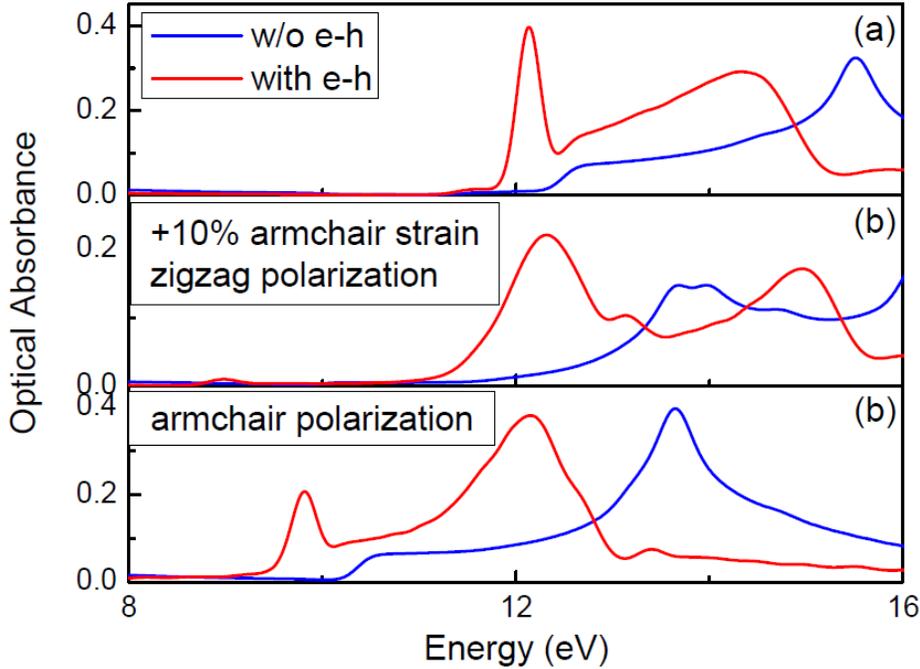

Figure 5. High-frequency optical absorption spectra of pristine (a) and armchair strained graphene with the incident light polarized along the zigzag direction (b) and armchair direction (c). A 0.15 eV Gaussian broadening is applied to obtain the smooth absorption curve. The bright exciton is marked by an arrow in (c).

For the armchair strained graphene, we observe the similar anisotropic optical absorption and enhanced excitonic effects as shown in Figure 5. The special character of armchair strained graphene is that we only find one bright nearly bound exciton for the incident light polarized along the armchair direction as shown in Figure 5 (c); for the zigzag polarization case (Figure 5 (b)), no nearly bound exciton (isolated peak) is identified and the whole spectrum is dominated by resonant excitonic effects.

**E. Exciton wave functions of strained graphene**

Since the applied strain significantly changes the geometry of graphene, it will inevitably modify electronic wave functions and associated excitons. In order to better understand excitonic effects on strained graphene, we have plotted wave functions of interested excitonic states.

According to the Tamm-Dancoff approximation,[30] the wave function of an exciton $S$ can be written in

$$\Phi_S(\vec{x}_e, \vec{x}_h) = \sum_k \sum_v^{hole} \sum_c^{elec} A_{vck}^S \phi_{ck}(\vec{x}_e) \phi_{vk}^*(\vec{x}_h), \qquad (2)$$

where the coefficient $A_{vck}^S$ is the solution to the BSE as shown in Eq (1). To visualize the *e-h* correlation of an excitonic state, we fix the position of the hole and plot the squared amplitude as function of the electron position. For clarity, we have integrated the wave function along the z direction (perpendicular to the graphene sheet) as

$$|\Phi_S(x_e, y_e, \vec{x}_h = 0)|^2 = \int dz_e |\Phi_S(\vec{x}_e, \vec{x}_h = 0)|^2. \qquad (3)$$

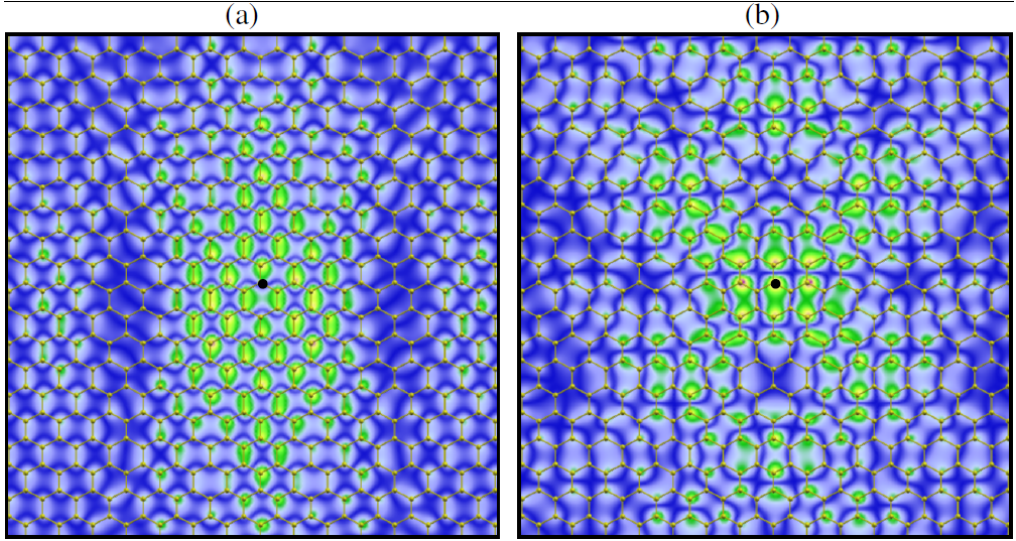

Figure 6. Wave function of bright resonant excitonic states in zigzag strained graphene. (a) a state with excitation energy of 2.84 eV, and (b) a state with excitation energy of 4.66 eV. Plotted is the squared electron amplitude given that the hole is fixed at the position marked with the black spot. The distributions have been integrated along the direction perpendicular to the plane of graphene.

Plotted in Figure 6 are two bright excitons from the two prominent absorption peaks (A and B) of zigzag strained graphene in Figure 2 (c). As we expect, these excitons are resonant states so that they have characteristic residual wave functions that are not decaying away from the hole. Moreover, because of the applied strain, those excitons plotted in Figure 6 (a) and (b) both display a significantly anisotropic distribution. In particular, the electronic wave function of the exciton plotted in Figure 6 (a) display a significant preference along the armchair direction.

Meanwhile, we plot the similar excitonic states of armchair strained graphene in Figure 7. In addition to the common features of Figures 6 and 7, such as the anisotropic distributions and resonant states, we find an interesting relation between these plot excitons. For example, the wave function in figure 6 (a) is similar to that in Figure 7 (b). The reason is from the fact that both these two excitons are consisting of similar wave functions, which are $\pi$ and $\pi^*$ wave functions at the $M'$ point of the first BZ. In Figures 1 (c) and (d), applied strain changes the energy of corresponding excitons, increasing the energy gap at the $M'$ point by armchair strain and decreasing it by zigzag strain. As a result, these similar excitons own different energy.

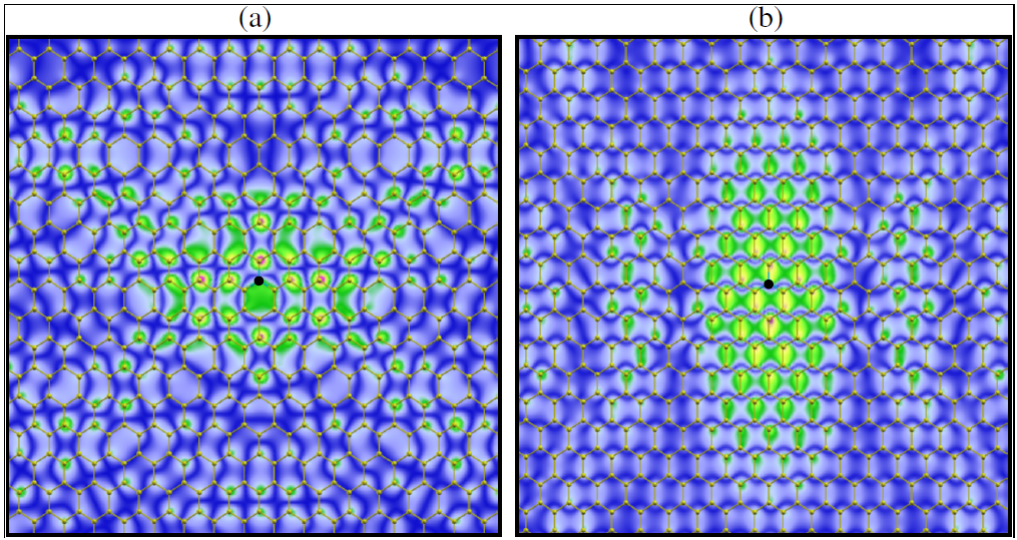

Figure 7. Wave function of bright resonant excitonic states in armchair strained graphene. (a) a state with excitation energy of 3.16 eV, and (b) a state with excitation energy of 4.79 eV. Plotted is the squared electron amplitude given that the hole is fixed at the position marked with the black spot.

## IV. CONCLUSIONS

In conclusion, the electronic band structure and optical absorption spectra of uniaxial strained graphene are studied by the first-principles GW+BSE approach. The anisotropic Fermi velocity is identified and the Dirac cone is shown to be tilted by applied strain. Consequently, the optical absorption spectra exhibit substantial strain effects: the optical absorption is anisotropic and strongly depending the polarization direction of the incident light and the strain direction; the characteristic single absorption peak from $\pi$ and $\pi^*$ transitions is split into two peaks and both display enhanced excitonic effects. At the same time, the high-frequency optical absorption involved with $\sigma$ and $\pi$ electronic states also display more new features than those of pristine graphene. All these new optical responses of strained graphene not only provide us efficient approaches to study the electronic structure and many-electron effects in such two-dimensional semimetal but also shed light on engineering novel optoelectronic applications of graphene.

## ACKNOWLEDGEMENT

Support from the International Center for Advanced Renewable Energy and Sustainability (I-CARES) in Washington University is gratefully acknowledged. We acknowledge computational resources support by the Lonestar of Teragrid at the Texas Advanced Computing Center (TACC) and the National Energy Research Scientific Computing Center (NERSC) supported by the U.S. Department of Energy. We also acknowledge the use the BerkeleyGW package.

## REFERENCES

1. K.S. Novoselov, A. K. Geim, S. V. Morozov, D. Jiang, Y. Zhang, S.V. Dubonos, I. V. Grigorieva, and A.A. Firsov: Electric field effect in atomically thin carbon films. *Science* **306**, 666 (2004).
2. K.S. Novoselov, A. K. Geim, S. V. Morozov, D. Jiang, M. I. Katsnelson, I. V. Grigorieva, and A.A. Firsov: Two-dimensional gas of massless Dirac fermions in graphene. *Nature* **438**, 197 (438).
3. Y. Zhang, Y.-W. Tan, H. L. Stormer, and P. Kim: Experimental observation of the quantum Hall effect and Berry's phase in graphene. *Nature* **438**, 201 (2005).
4. C. Berger, Z. Song, T. Li, X. Li, A. Y. Ogbazghi, R. Feng, Z. Dai, A. N. Marchenkov, E. H. Conrad, P. N. First, and W. A. de Heer: Ultrathin epitaxial graphite: 2D electron gas properties and a route toward graphene-based nanoelectronics. *J. Phys. Chem. B* **108**, 19912 (2004).
5. A.K. Geim and K.S. Novoselov: The rise of graphene, *Nature Mater.* **6**, 183 (2007).
6. A.H. Castro Neto, F. Guinea, N.M.R. Peres, K.S. Novoselov and A.K. Geim: The electronic properties of graphene. *Rev. Mod. Phys.* **81**, 109 (2009).
7. V.P. Gusynin and S.G. Sharapov: Transport of Dirac quasiparticles in graphene: Hall and optical conductivities. *Phys. Rev. B* **73**, 245411 (2006).
8. R.R. Nair, P. Blake, A.N. Grigorenko, K.S. Novoselov, T.J. Booth, T. Stauber, N.M.R. Peres and A.K. Geim: Fine structure constant defines visual transparency of graphene. *Science* **320**, 1308 (2008).
9. V.G. Kravets, A.N. Grigorenko, R.R. Nair, P. Blake, S. Anissimova, K.S. Novoselov, A.K. Geim: Spectroscopic ellipsometry of graphene and an exciton-shifted van Hove peak in absorption. *Phys. Rev. B* **81**, 155413 (2010).
10. K.F. Mak, M.Y. Sfeir, Y. Wu, H. Lui, J.A. Misewich, and T.F. Heinz: Measurement of the optical conductivity of graphene. *Phys. Rev. Lett.* **101**, 196405 (2008).
11. Feng Wang, Yuanbo Zhang, Chuanshan Tian, Caglar Girit, Alex Zettl, Michael Crommie, and Y. Ron Shen: Gate-variable optical transitions in graphene. *Science*

**320**, 206 (2008).
12. K.S. Kim, Y. Zhao, H. Jang, S.Y. Lee, J.M. Kim, K.S. Kim, J. H. Ahn, P. Kim, J. Choi, and B.H. Hong: Large-scale pattern growth of graphene films for stretchable transparent electrodes. *Nature* **457**, 706 (2009).
13. C. Lee, X. Wei, J.W. Kysar, and J. Hone: Measurement of the elastic properties and intrinsic strength of monolayer graphene. *Science* **321**, 385 (2008).
14. F.M.D. Pellegrino, G.G. N. Angilella, and R. Pucci: Effect of uniaxial strain on the reflectivity of graphene. *High Press. Res.* **29** 569 (2009).
15. F.M.D. Pellegrino, G.G.N. Angilella, and R. Pucci: Strain effect on the optical conductivity of graphene. *Phys. Rev. B* **81**, 035411 (2010).
16. A. Sinner, A. Sedrakyan and K. Ziegler: Optical conductivity of graphene in the presence of random lattice deformations. *Phys. Rev. B* **83**, 155115 (2011).
17. V.M. Pereira, R.M. Ribeiro, N.M.R. Peres, and A.H. Castro Neto: Optical properties of strained graphene. *EPL* **92**, 67001 (2011).
18. L. Yang, J. Deslippe, C.-H. Park, M.L. Cohen, and S.G. Louie: Excitonic effects on the optical response of graphene and bilayer graphene. *Phys. Rev. Lett.* **103**, 186802 (2009).
19. K.F. Mak, J. Shan, and T.F. Heinz: Seeing many-body effects in single- and few-Layer graphene: observation of two-dimensional saddle-Point excitons. *Phys. Rev. Lett.* **106**, 046401 (2011).
20. L. Yang: Excitonic Effects on Optical Absorption Spectra of Doped Graphene. *Nano Lett.*, Article ASAP (2011).
21. M. Rohlfing and S.G. Louie: Electron-hole excitations and optical spectra from first principles. *Phys. Rev. B* **62**, 4927 (2000).
22. P. Hohenberg and W. Kohn: Inhomogeneous electron gas. *Phys. Rev.* **136**, B864 (1964).
23. W. Kohn and L.J. Sham: Self-consistent equations including exchange and correlation effects. *Phys. Rev.* **140**, A1133 (1965).
24. L. Hedin: New method for calculating the one-particle Green's function with application to the electron-Gas problem. *Phys. Rev.* **139**, A796 (1965).
25. M.S. Hybertsen and S.G. Louie: Electron correlation in semiconductors and insulators: Band gaps and quasiparticle energies. Phys. Rev. B 34, 5390 (1986).
26. G. Onida, L. Reining, and A. Rubio: Electronic excitations: density-functional versus many-body Green's-function approaches. *Rev. Mod. Phys.* **74**, 601 (2002).
27. N. Troullier and J.L. Martins: Efficient pseudopotentials for plane-wave calculations. *Phys. Rev. B* **43**, 1993 (1991).
28. P. E. Trevisanutto, M. Holzmann, M. Cote, and V. Olevano: Ab initio high-energy excitonic effects in graphite and graphene. *Phys. Rev. B* **81**, 121405 (2010).
29. L. Yang: Excitons in intrinsic and bilayer graphene. *Phys. Rev. B* **83**, 085405 (2011).
30. A. Fetter and J.D. Walecka: Quantum Theory of Many Particle Systems (McGraw-Hill, San Francisco, 1971), p. 538.